\documentclass[aps,prl,letterpaper,superscriptaddress,10pt,twocolumn,floatfix]{revtex4-1}
\usepackage[colorlinks=true,allcolors=blue]{hyperref}
\usepackage{amssymb}
\usepackage{amsmath}
\usepackage{float}
\usepackage{graphicx}
\usepackage[caption=false]{subfig}
\usepackage{amsfonts}
\usepackage{xcolor}
\usepackage{epsfig}
\usepackage{color}
\usepackage{bm}
\usepackage{tabularx}
\usepackage{multirow}
\usepackage{mathtools}
\usepackage{xfrac}
\usepackage{blkarray}
\usepackage{bbold}
\usepackage[mathscr]{euscript}
\usepackage{autobreak}
\usepackage{comment}
\usepackage{makecell}
\usepackage{soul, color, xcolor}
\soulregister{\cite}7 
\soulregister{\citep}7 
\soulregister{\citet}7
\soulregister{\ref}7
\soulregister{\onlinecite}7
\usepackage{float}
\usepackage{amsmath}
\makeatletter

\newcommand{\Rmnum}[1]{\expandafter\@slowromancap\romannumeral #1@}
\makeatother

\begin{document}
\title{Solid state defect emitters with no electrical activity}

\author{Pei Li}
\affiliation{School of Integrated Circuit Science and Engineering, Tianjin University of Technology, Tianjin 300384, China}
\affiliation{Beijing Computational Science Research Center, Beijing 100193, China}
\affiliation{HUN-REN Wigner Research Centre for Physics, P.O.\ Box 49, H-1525 Budapest, Hungary}

\author{Song Li}
\affiliation{HUN-REN Wigner Research Centre for Physics, P.O.\ Box 49, H-1525 Budapest, Hungary}

\author{P\'eter Udvarhelyi}
\affiliation{HUN-REN Wigner Research Centre for Physics, P.O.\ Box 49, H-1525 Budapest, Hungary}
\affiliation{Department of Atomic Physics, Institute of Physics, Budapest University of Technology and Economics, M\H{u}egyetem rakpart 3., H-1111 Budapest, Hungary}

\author{Bing Huang}
\affiliation{Beijing Computational Science Research Center, Beijing 100193, China}
\affiliation{Department of Physics, Beijing Normal University, Beijing, 100875, China}

\author{Adam Gali}
\affiliation{HUN-REN Wigner Research Centre for Physics, P.O.\ Box 49, H-1525 Budapest, Hungary}
\affiliation{Department of Atomic Physics, Institute of Physics, Budapest University of Technology and Economics, M\H{u}egyetem rakpart 3., H-1111 Budapest, Hungary}

\date{\today}

\begin{abstract}
Point defects may introduce defect levels into the fundamental band gap of the host semiconductors that alter the electrical properties of the material. As a consequence, the in-gap defect levels and states automatically lower the threshold energy of optical excitation associated with the optical gap of the host semiconductor. It is, therefore, a common assumption that solid state defect emitters in semiconductors ultimately alter the conductivity of the host. Here we demonstrate on a particular defect in 4H silicon carbide that a yet unrecognized class of point defects exists which are optically active but electrically inactive in the ground state.
\end{abstract}

\maketitle


Point defects in semiconductors play a pivotal role in determining the electrical and optical properties of the host material. Understanding the physical fundaments of point defects in semiconductors was a key to arrive at the concept of opto-electronics devices, photovoltaics and energy storage devices, and very recently, state-of-the-art quantum information processing realisations~\citep{polman2016photovoltaic, awschalom2013quantum, zwanenburg2013silicon, awschalom2018quantum, wolfowicz2021quantum} which have been shaped the socio-economical environment at global scale. Point defects may introduce defect levels (DLs) within the host semiconductor's fundamental band gap, thereby influencing its electrical conductivity, i.e., electrically active point defects~\citep{queisser1998defects, alkauskas2016tutorial}. Notably, these in-gap DLs and associated states also impact the optical properties of the material by reducing the optical excitation threshold energy compared to the perfect semiconductor's optical gap~\citep{feil2023electrically, redjem2020single, li2022carbon, Gali2023}. As a consequence, a common assumption is that solid-state defect emitters modify the host material's electrical conductivity~\citep{tsai2022antisite, ping2021computational, gordon2013quantum, weber2010quantum}. We show below that this common assumption is not generally valid. 

\begin{figure}[htb]
\includegraphics[width=0.4\textwidth]{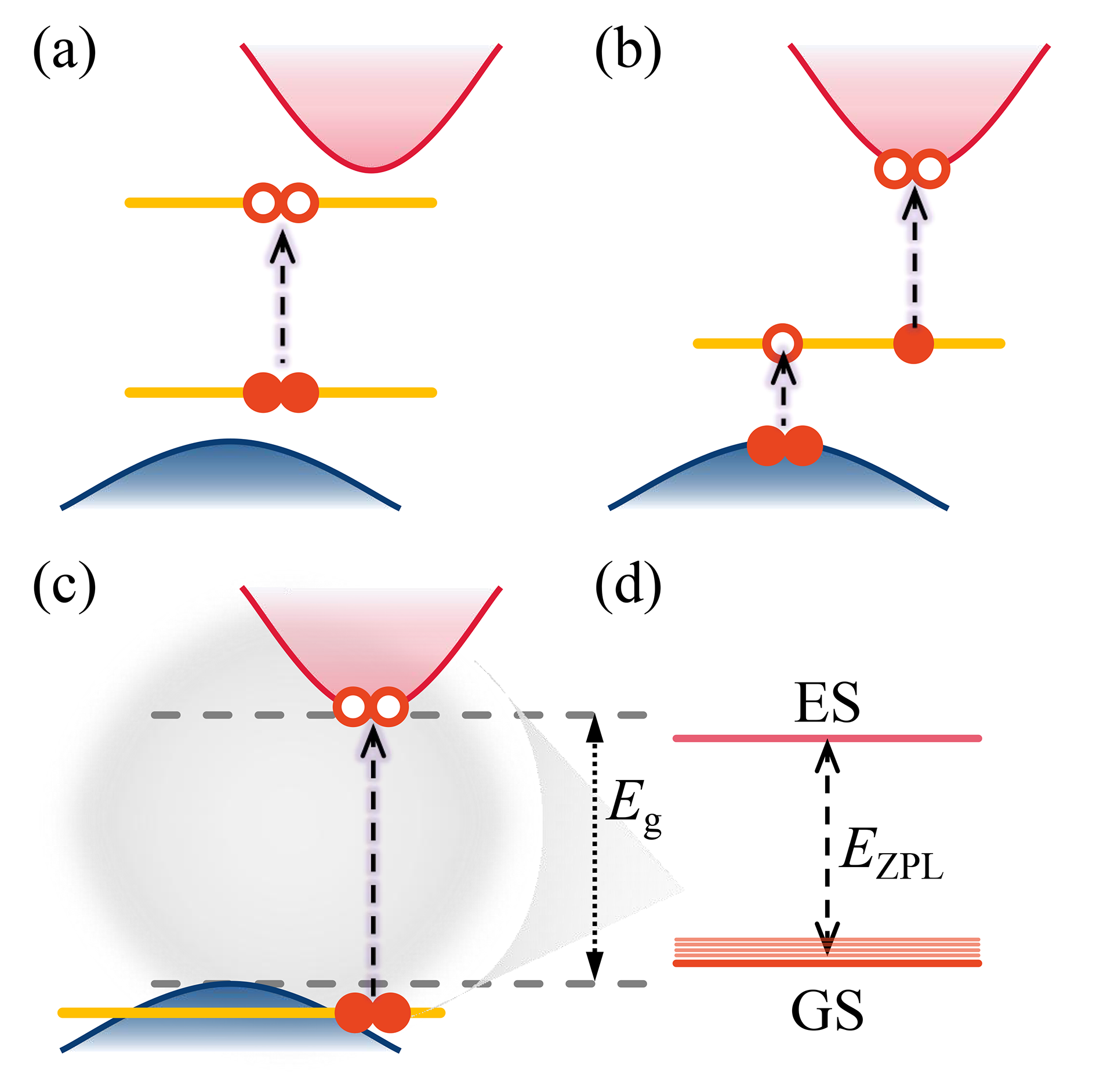}
\caption{\label{Figure_1}%
The possible optical transition mechanisms within indirect band gap semiconductors. (a) Occupied and unoccupied deep DLs. (b) Deep DL and band edge. (c) Electrically inactive DL and band edge: establishing EIDE. (d) Comparison between the values of zero-phonon-line ($E_\text{ZPL}$) and band gap ($E_\text{g}$) for the defect category in (c). $E_\text{ZPL}$ is the energy difference between the optical excited (ES) and ground state (GS) of optical transitions with no-phonon whereas the thinner lines are the optical transition energies when phonons are participating in the optical transition.} 
\end{figure}

In Fig.~\ref{Figure_1}, we depict the possible optical transition mechanisms within semiconductors. Many point defects introduce multiple deep DLs into the fundamental band gap that could dramatically modify the electrical properties of the host because these deep levels often participate in carrier trapping and recombination events. 
In these defects, optical transition could occur between the occupied and unoccupied DLs in the gap 
[see Fig.~\ref{Figure_1}(a)]. Alternatively, the optical transition can occur between localized DLs and the band edge, either valence band maximum (VBM) or conduction band minimum (CBM) (e.g. Ref.~\citep{li2023carbon}). The respective excited states may be called pseudo-acceptor and pseudo-donor states as they show a Rydberg-series of the excited states converging towards the ionization threshold [see Fig.~\ref{Figure_1}(b)]. We note that the optical excitation threshold energies are lower than the electrical gap between the occupied and unoccupied states participating in the optical transition because of the attracting electron-hole interaction in the excited state. By harnessing this excitonic effect, we suggest a category of point defects that act as emitters and are electrically inactive at the same time. A defect may introduce just one occupied DL below VBM without disturbing the bands close to VBM, so the defect is electrically inactive in the ground state and its positive charge state is not stable. This defect can be optically excited where the hole is localized in the resonant DL whereas the electron occupies a state split from CBM that builds up a pseudo-donor excited state. The exciton binding energy in the excited state could shift the excitation energy below the optical band gap of the host semiconductor with establishing a solid state defect emitter [see Fig.~\ref{Figure_1}(c)]. We label these defects as EIDE after the expression of electrically inactive defect emitters in the context.  

In this study, we demonstrate the principles of EIDE on a tri-carbon interstitial cluster in 4H silicon carbide (SiC) which produces an ultraviolet emission below the gap of 4H-SiC. We show by first principles calculations that the given defect has zero-phonon-line (ZPL) emission with characteristic local vibration modes in the photoluminescence (PL) spectrum which agrees well with a previously reported defect emitter, the so-called D$_\text{II}$ center in 4H-SiC~\citep{sridhara1998, steeds2006origin, sullivan2007study}. The optical excited state is a pseudo-donor type and it does not show electrical activity. The effect is mediated by the attractive electron-hole interaction in the optical excited state of the defect enhanced by the resonant defect states which is strikingly paramount in indirect semiconductors. This example unveils the EIDE category of point defects in solids. We discuss potential dopants to engineer such defects in semiconductors.

\section{Results}
\label{sec:results}

4H-SiC is a wide band gap indirect semiconductor which is a platform for high-power, high temperature electronic devices~\citep{pearton2013processing, lutz2011semiconductor, shur2006sic} as well as quantum information processing~\citep{SiC-review-paper-Son-Awschalom, SiC-csore-review-paper-for-defects} that makes it unique among the technological mature semiconductors. The band gap of 4H-SiC is slightly reduced at elevated temperatures~(see Ref.~\citep{cannuccia2020thermal} and references therein). The VBM and CBM are located at $\Gamma$-point and $M$-point, respectively. The excitonic band gap of the material is 3.265~eV at 2~K~\citep{choyke1964}, where the crystal phonon replica dominates the PL spectrum upon above-band-gap illumination; nevertheless, the ZPL of the free exciton can be weakly observed too because the free exciton may gain some momentum from the defects in 4H-SiC~\citep{ivanov1998phonon}. The PL spectrum of the bound exciton of the shallow donors (nitrogen substituting carbon in the lattice at the so-called quasicubic and hexagonal sites) can be well observed at 2~K, where the respective ZPL emissions at 3.243~eV and 3.256~eV are more pronounced as the defect potential breaks the translational symmetry of the crystal. Nevertheless, the phonon sideband still dominates in the respective PL spectrum as expected for an indirect semiconductor~\citep{ivanov1998phonon}.

In the followings, we focus our attention to the ultraviolet D$_\text{II}$ color center in 4H-SiC recorded near cryogenic temperatures~\citep{sullivan2007study}. The D$_\text{II}$ shows up a sharp ZPL at 3.20~eV which is only $\approx 60$~meV lower than that of the free exciton in 4H-SiC~\citep{10.1063/1.113205, klahold2018high, ivanov1998phonon}. A characteristic phonon sideband was also observed in the D$_\text{II}$ PL spectrum with sharp features that were associated with local vibration modes (LVMs) of the underlying defect~\citep{sullivan2007study}. We note that various reports on the D$_\text{II}$ PL spectrum exhibit different numbers of sharp features associated with LVMs~\citep{sridhara1998, sullivan2007study}. No single defect observation has yet been carried out for D$_\text{II}$ center~\citep{PhysRevB.69.235202, jiang2012carbon, MATTAUSCH2001656, patrick1973localized}, thus some sharp features may not belong to the defect and could overlap with the PL spectrum of other defects. Nevertheless, previous studies attempted to identify this color center by calculating the LVMs of the defect models and compare those to the observed features in the phonon sideband~\cite{MATTAUSCH2001656, gali2003aggregation, jiang2012carbon}. The most recent study proposed the tri-carbon interstitial cluster as the origin of the D$_\text{II}$ center which was corroborated by molecular dynamics calculations with revealing the high-temperature stability of the defect in line with the observations~\citep{jiang2012carbon}. However, the nature of optical transition has not yet been discussed for any models. Furthermore, it has been recently shown that not all the calculated LVMs show up in the observed phonon sideband of the PL spectrum of other carbon clusters in 4H-SiC~\citep{li2023carbon}, thus identification of this ultraviolet color center has not yet been established.    

\begin{figure}[htb]
\includegraphics[width=0.4\textwidth]{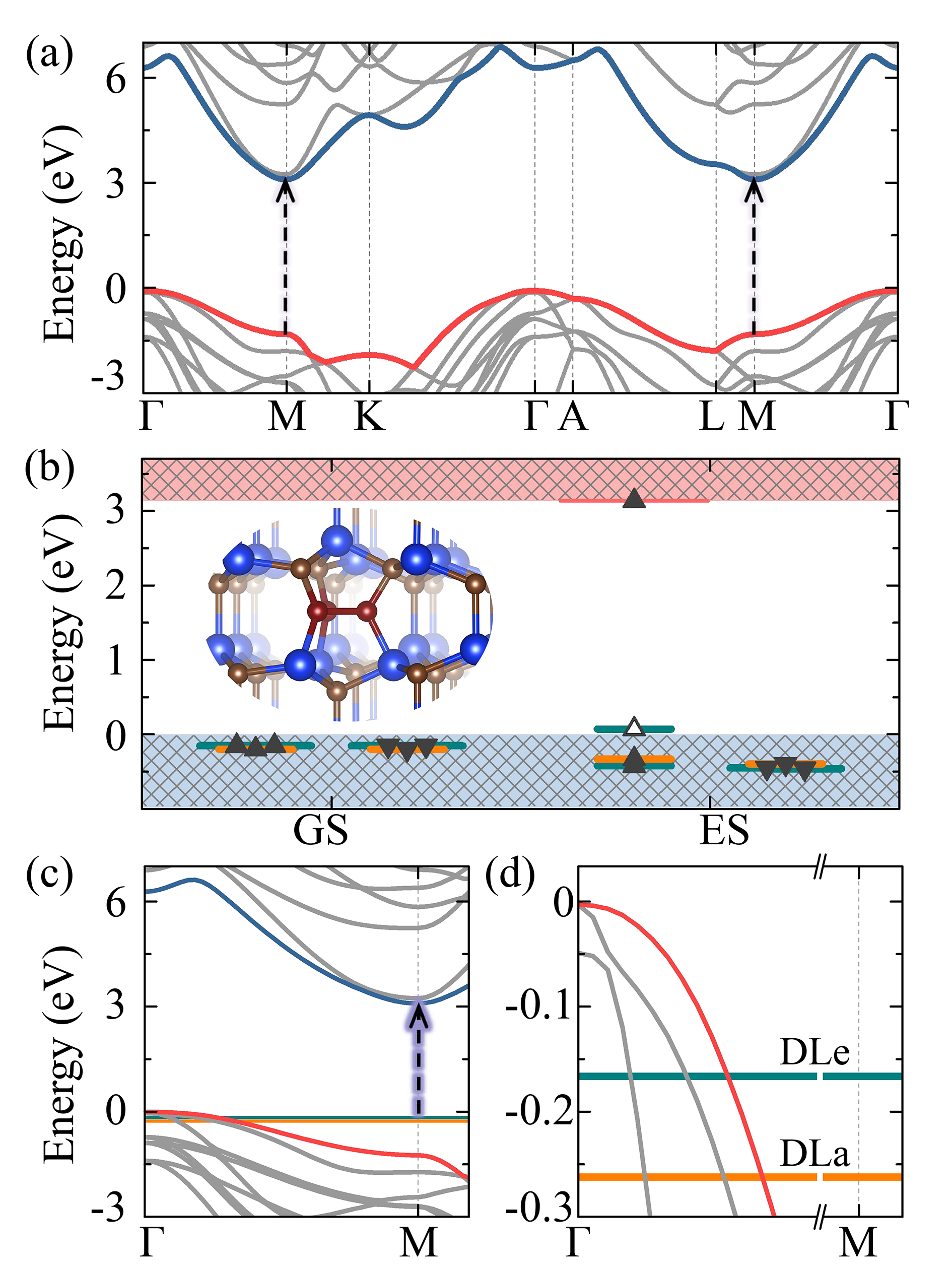}
\caption{\label{Figure_2}%
(a) The band structure of 4H-SiC. (b) The calculated Kohn-Sham DLs of the ground state (GS) and excited state (ES) for tri-carbon interstitial cluster. The occupied and unoccupied DLs are labelled by filled and empty triangles, respectively. Inset: the configuration of tri-carbon interstitial cluster. (c) The band structure of 4H-SiC with tri-carbon interstitial cluster. (d) The local enlarged drawing of (c). The degenerate $e$ states (DLe) and $a_1$ state (DLa) are labelled in green and orange, respectively.}
\end{figure}

We use the tri-carbon interstitial cluster as a working model for the D$_\text{II}$ PL center in 4H-SiC. First, the electronic structure and possible optical transitions of the host semiconductor is analyzed. The band structure of 4H-SiC is depicted in Fig.~\ref{Figure_2}(a). Our HSE06 calculations yield 3.17~eV electronic band gap for the host 4H-SiC which underestimates the low-temperature data by about 0.1~eV. The electron-hole pair in a free exciton has a crystal momentum $\text{k}_M$, where $\text{k}_M$ is the momentum corresponding to the $M$-point. Consequently, direct recombination of the electron at VBM and the hole at CBM is forbidden because of the crystal-momentum conservation law. To conserve the crystal momentum the free exciton recombination is only possible with the assistance of another particle (or quasiparticle). Therefore, the minimum direct optical transition in 4H-SiC occurs between the band edges at the $M$-point, with a gap of approximately 4.41~eV, and is larger than the indirect band gap.

The tri-carbon interstitial cluster is composed of three carbon interstitials bridging three adjacent on-axis Si-C bonds at the hexagonal site [see Fig.~\ref{Figure_2}(b)], which is the most stable among all possible configurations~\citep{li2023carbon}. The calculated Kohn-Sham DLs for the neutral tri-carbon interstitial cluster are shown in Fig.~\ref{Figure_2}(b). In the ground state, a double degenerate $e$ state (VBM$-$0.15~eV) and an $a_1$ state (VBM$-$0.20 eV) appear that show no dispersion unlike the host bands. We find also resonant states in this energy region when crossed with the host bands that show up quasilocalization but heavily mixed with the host bands. No further DLs emerge in the fundamental band gap and the VBM is basically unaffected by the presence of DLs (see Supplementary Fig.~1), implying that the defect is electrical inactive [see Figs.~\ref{Figure_2}(c,d)]. Indeed, we find in our density functional theory (DFT) calculations that the positively charged defect is not stable (see Supplementary Fig.~2).

\begin{figure}[htb]
\includegraphics[width=0.4\textwidth]{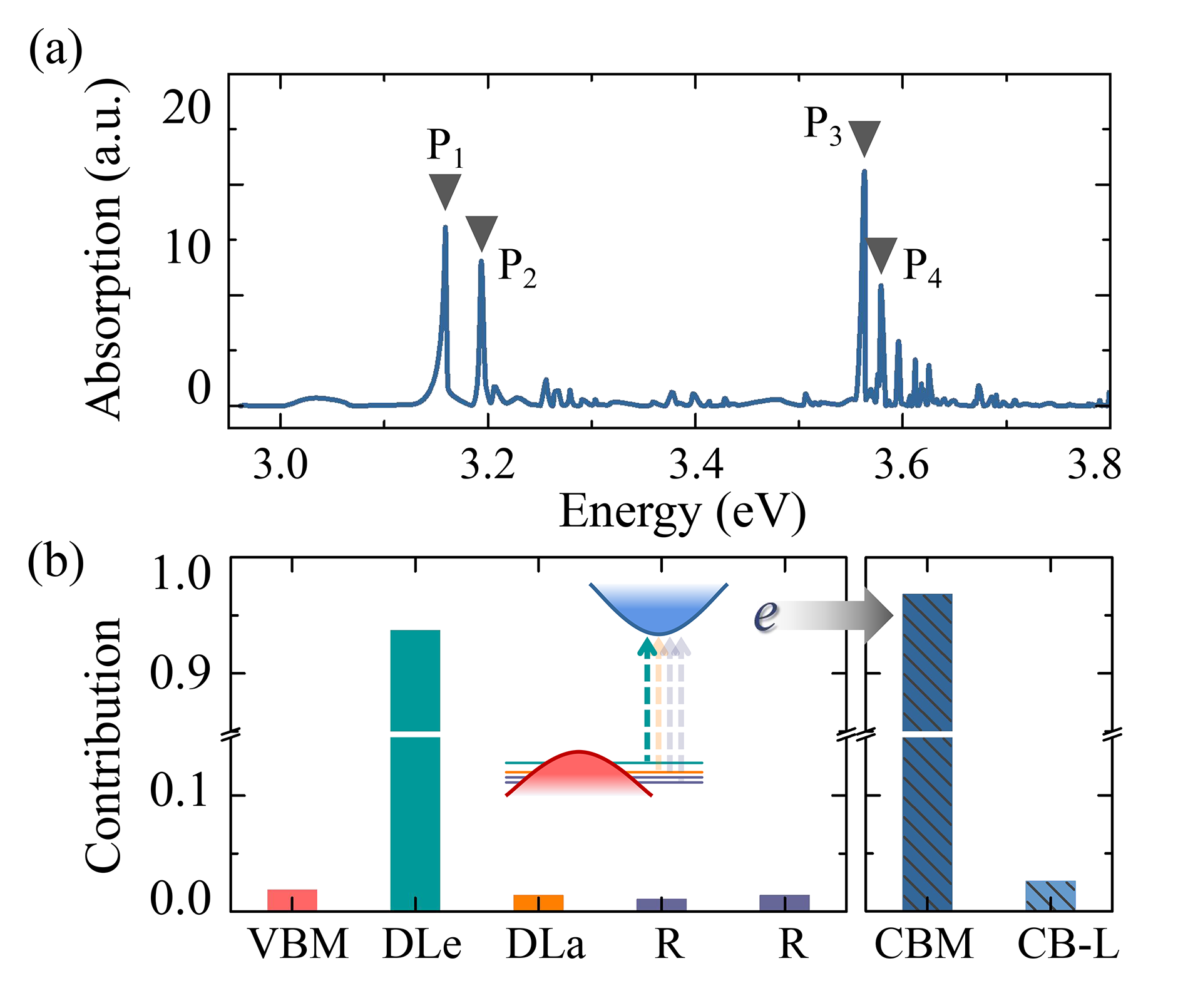}
\caption{\label{Figure_3}%
(a) The optical transition intensity calculated within BSE. (b) The contributions of excitations originating from VBM, DLs, and resonances state (R) to CBM and conduction band in the $L$-point (CB-L) for the peak $\text{P}_\text{1}$ in (a). Inset: The diagram of excitation.}
\end{figure}

It can be recognized that the DLs will be well isolated from the bands around the $M$-point and it may be expected that the electron from the DLs can be photoexcited to the CBM that constitutes a pseudo-donor excitonic state. We first studied the nature of the exciton and the oscillator strength of the optical transitions by a many-body perturbation method (see Methods) with that the excitonic effects can be accurately calculated [see Fig.~\ref{Figure_3}(a)]. Here we focus to the lowest-energy bright transitions that only play the role in the PL emission of the defect.  
Bright transitions occur at around 3.2~eV that we label by $\text{P}_\text{1}$ and $\text{P}_\text{2}$ in Fig.~\ref{Figure_3}(a). In both peaks the hole part of the exciton wavefunction is dominantly built up from the DLs and the electron part is located in the conduction bands at the $M$-point. The contribution is about 95\% and 99\% for $\text{P}_\text{1}$ and $\text{P}_\text{2}$ peaks, respectively. In contrast to $\text{P}_\text{1}$ and $\text{P}_\text{2}$ peaks, the $\text{P}_\text{3}$ and $\text{P}_\text{4}$ peaks in Fig.~\ref{Figure_3}(a) are mainly caused by the excitation from DLs to the conduction bands at the $L$-point, collectively constituting approximately 87\% and 82\% of the total intensity. We here analyze the origin of the $\text{P}_\text{1}$ peak in detail whereas the analysis of the other peaks can be found in Supplementary Note 3. The calculated many-body perturbation theory band gap between the double degenerate $e$ state (DLe) and the CBM at the $M$-point is 3.57~eV, thus the exciton binding energy is about 0.4~eV. The contribution of the electron-hole pairs to the $\text{P}_\text{1}$ exciton is depicted in Fig.~\ref{Figure_3}(b). The foremost contribution arises from excitation originating from the degenerate DLe to the CBM, constituting approximately 93.71\% of the total intensity. The excitation from the $a$ state (DLa) to CBM contributes by about 1.40\%. Thus, it can be concluded that the vast majority of the hole wavefunction has a localized nature whereas the electron wavefunction is from the CBM in the low-energy bound exciton.

Based on our findings about the lowest energy bright excited state in the BSE spectrum, we employed HSE06 $\Delta$SCF method where an electron was promoted from the DLe to the CBM, in order to take into account the geometry relaxation in the optical excited state. This calculation makes it possible to simulate the ZPL energy and the phonon sideband in the PL spectrum. We assume that the vertical excitation energy can be well calculated by the many-body perturbation method and then we apply finite size error corrections of the supercell method, and then the reorganization energy is subtracted as obtained by the HSE06 $\Delta$SCF method to arrive at the final ZPL energy. We note that the $\Delta$SCF procedure will self-consistently change the Kohn-Sham states and levels accordingly. In the excited state, the hole DL pops up in the band gap close to VBM by 0.07~eV [see Fig.~\ref{Figure_2}(b)]. The redistribution in the localization of the defect wavefunctions results in relatively high forces on the ions and the defect reconstructs in the optical excited state with a reorganization energy of about 0.13~eV with slightly reducing the symmetry of the defect. We considered the P$_1$ peak in the BSE spectrum as the vertical excitation energy for which we applied finite size error corrections (adding 0.10~eV) that account for the pseudo-donor nature of the excited state as explained in Supplementary Note 4. The final simulated ZPL is at 3.13~eV, in good agreement with experimental data. 

We then computed the phonon sideband of the PL spectrum within Huang-Rhys theory with using the excited state's geometry as obtained by HSE06 $\Delta$SCF method~\citep{huang1950theory, alkauskas2014first, PhysRevLett.103.186404}. The PL spectrum of the tri-carbon interstitial cluster is shown in Fig.~\ref{Figure_4} and the respective vibration frequencies are listed in Supplementary Table I. Because of the symmetry reduction, both symmetry-breaking $E$ and symmetry conserving $A_1$ vibration modes participate to the phonon sideband of the PL spectrum. The double degenerate ($E$-mode) highest LVMs at 162~meV are stretching modes from two of the three approximately vertical C-C bonds. The third LVM is a stretching mode of the three C-C bonds, which preserves the symmetry of the defect ($A_1$-mode). The fourth LVM is a breathing mode of the triangle formed by the three carbon interstitials. The respective fifth and sixth LVMs result from the axial vibration of the carbon atoms right below and above the center of the tri-carbon interstitial cluster. We conclude that the calculated ZPL energy and the LVM structure well agree with those of the D$_\text{II}$ color center in 4H-SiC (see Supplementary Note 5).   

\begin{figure}[htb]
\includegraphics[width=0.4\textwidth]{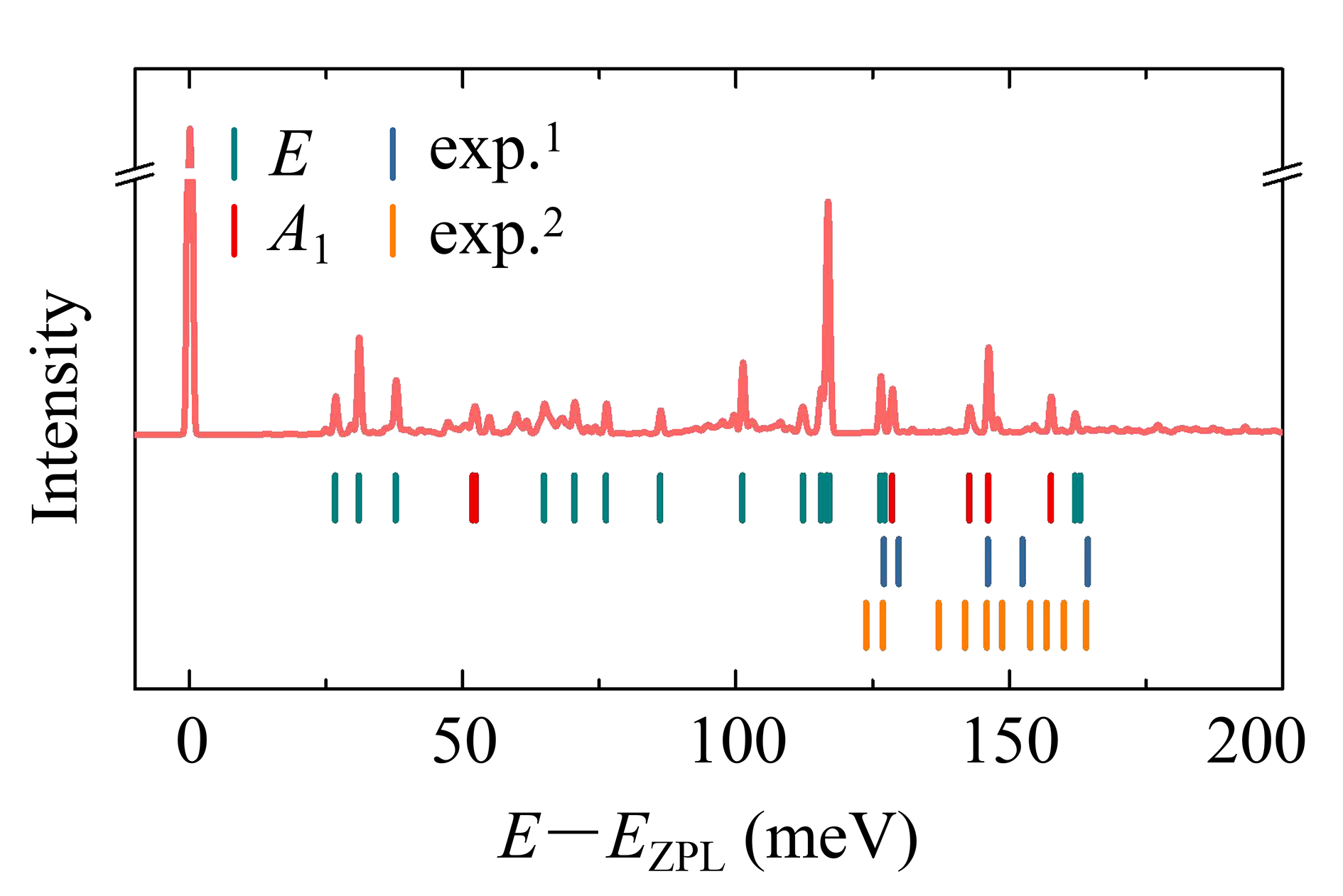}
\caption{\label{Figure_4}%
The PL spectrum of the tri-carbon interstitial cluster. The $E$ and $A_\text{1}$ modes are marked with green and red vertical lines below the corresponding LVMs. The LVMs (blue~\citep{sridhara1998} and orange~\citep{sullivan2007study} vertical lines) of $\text{D}_\text{II}$ PL spectrum with frequency higher than that of 4H-SiC bulk phonon spectrum are also listed for comparison.}
\end{figure}


\section{Discussion}
\label{sec:discussion}

We demonstrated that the D$_\text{II}$ color center is a point defect which belongs to the EIDE category. In this particular case, the excited state has a pseudo-donor nature. This explains the fact that the ZPL energy of the D$_\text{II}$ center scales with the widths of the band gap of SiC polytypes, c.f., Refs.~\citep{patrick1973localized, sullivan2007study, sridhara1998}. One can generalize the feasible electronic structure of EIDE that (i) either an occupied DL should lie below but not too far from VBM but should be deep enough not to realize a stable positive charge state (ii) or an empty DL should lie above but not too far from CBM but high enough not to realize a stable negative charge state or (iii) the combination of both. The bound exciton state may be easier realized in an indirect than a direct semiconductor where the attractive electron-hole interaction of the bound exciton and the geometry relaxation in the excited state could result in a ZPL emission below the optical gap of the host semiconductor. The wavelength of these emitters depend on the fundamental band gap the host semiconductors and the strength of exciton binding energies in the excited of the defect. The last term depends on the screening effects of the material. Semiconductors with moderate and low screening can host defects with exciton binding energies that may exceed hundreds of millielectronvolts which establishes a considerable playground to realize EIDEs. 

Now one may ask how one can engineer EIDEs in various materials. Electrically inactive defect in semiconductors strongly implies that (i) vacancy or vacancy complexes can be disregarded as dangling bonds introduce levels in the fundamental band gap, (ii) the substitutional defect or dopant should be isovalent with the host atom to avoid donating or accepting electrons to the crystal, and (iii) interstitial or interstitial cluster defects should introduce even number of electrons to realize a closed-shell singlet ground state and they should possess stronger chemical bonds than those constituting of the host crystal so the occupied and unoccupied defect levels would presumably lie outside of the fundamental band gap. Certainly, the EIDE defect models should be checked one-by-one in a given semiconductor. This is not surprising as this is the case, e.g., for the effective mass donor (n-type doping) or acceptor (p-type doping) models. Although, the usual recipe for realizing donor and acceptor states is to substitute the host atom with a dopant possessing one extra and one less electron, respectively, this recipe does not always work. As an example, the nitrogen has five valence electrons that can replace silicon atom with four valence electrons in the silicon crystal; however, one chemical bond will be broken in the defect and the silicon dangling bond produces a deep level instead of the shallow effective mass donor. On the other hand, boron atom with three valence electrons substituting silicon acts as an effective mass acceptor in silicon crystal despite the fact that boron and nitrogen atoms sit in the same row of the periodic table. On the contrary, nitrogen substituting the carbon in SiC indeed acts as effective mass donor as was anticipated earlier in this paper. These examples highlight the quest of accurate \emph{ab initio} predictions of defect states because simple rules cannot be trustworthy applied for cases that are thought to be easy such as defect engineering of donors and acceptors in semiconductors. This study rather establishes the search criteria for electrically inactive defect emitters in semiconductors as quoted above. Recent advances in the theoretical spectroscopy of defects in semiconductors~\citep{Gali2023} make it possible to systematically search for defects with target properties that may lead to discoveries of EIDEs, n-type and p-type dopants or defect qubits in semiconductors. 

This study uncovers a class of defects that decouple optical and electrical effects. These findings propose an unexplored avenue for engineering semiconductor devices, suggesting the feasibility of creating defect species that offer independent control over optical and electrical functionalities within the same platform. The implications of this work extend to the design and optimization of highly integrated and miniaturized semiconductor devices that leverage both optical and electrical attributes for enhanced functionality. Because the modularity of the opto-electronics devices can be simplified this may lead to reduced cost of production and to contribution to green technology with lowering the energy consumption of the operation of these devices.

\section{Conclusion}
\label{sec:conclusion}

This study has revealed a phenomenon that expands our understanding of the interplay between defect-induced optical and electrical effects that was exemplified on a tri-carbon intersitial defect in 4H-SiC. Unlike the conventional assumption that solid-state defect emitters in semiconductors necessarily affect the electrical conductivity of the host, we demonstrate the existence of defects that introduce optically active but electrically inactive states. This phenomenon may be observed in rather indirect than direct band gap semiconductors where all the defect levels lie outside to the fundamental band gap but reside very close to the band edges that may alter the minimal optical excitation energy without influencing the electrical conductivity. By exploiting the attractive interaction between electrons and holes in the optical excited state, these defects can show significant optical activity without exerting any discernible impact on the electrical conductivity. This finding might pave the ways to realize new generation opto-electronics devices.

\section{Methods}
\label{sec:method}
All the first-principles calculations are performed using density functional theory (DFT) within the projector augmented wave potential plane-wave method, as implemented in the Vienna \textit{ab initio} simulation package (VASP)~\citep{PhysRevB.54.11169} with the projector augmented wave method~\citep{PhysRevB.50.17953}. The electron wave functions are expanded in plane-wave basis set limited by a cutoff of 420~eV. The fully relaxed geometries were obtained by minimizing the quantum mechanical forces between the ions falling below the threshold of 0.01~eV/\AA\ and the self-consistent calculations are converged to $10^{-5}$~eV.

The screened hybrid density functional of Heyd, Scuseria, and Ernzerhof (HSE06)~\citep{heyd2003hybrid} is employed to calculate the electronic structure. In this approach, we could mix part of nonlocal Hartree–Fock exchange to the generalized gradient approximation of Perdew, Burke, and Ernzerhof (PBE)~\citep{PhysRevLett.77.3865} with the default fraction ($\alpha$ = 0.25) and the inverse screening length at 0.2~\AA$^{-1}$. The calculated band gap is 3.17~eV. We embedded the tri-carbon interstitial cluster in a 576-atom 4H-SiC supercell which is sufficient to minimize the periodic defect-defect interaction. The single $\Gamma$-point scheme is convergent for the k-point sampling of the Brillouin zone (BZ). The $M$-point and $L$-point in the BZ are projected into the $\Gamma$-point in this supercell due to band folding where the lowest energy conduction bands occur in these k-points. The excited states were calculated by $\Delta$SCF method~\citep{PhysRevLett.103.186404}. We note that the reorganization energy and the optimized geometry of the optical excited state can be calculated with $\Delta$SCF method which are both important to predict the photoluminescence spectrum including the phonon sideband.

For the phonon modes, we calculated the corresponding dynamical matrix containing the second-order derivatives of the total energy by means of the PBE~\citep{PhysRevLett.77.3865} functional where all the atoms in the supercell were enabled to vibrate. In this case, we apply strict threshold parameters for the convergence of the electronic structure ($10^{-6}$~eV) and atomic forces ($10^{-3}$~eV/\AA ) in the geometry optimization procedure. These vibration modes are applied together with the HSE06 ground state and excited state geometries to simulate the PL spectrum of the given defects within Huang-Rhys theory~\citep{alkauskas2014first, PhysRevLett.103.186404}. This strategy worked well for deep defects in diamond (e.g., Refs.~\onlinecite{ThieringNV2017, ThieringG4V2018}).

To accurately consider the excitonic effect, many-body perturbation theory based on GW approximation with Bethe-Salpeter equation (BSE)~\citep{shishkin2006implementation, shishkin2007accurate} are used. We here use the supercell with 576-atom in order to reach close-to-converged calculation for the GW method which is very computationally demanding in the VASP implementation as more than 9000 bands were included in the single-shot $\text{G}_0\text{W}_0$ calculation. We note that the CBM at the $M$-point and the conduction band at the $L$-point are projected into the $\Gamma$-point for this supercell too. The energy cutoff for the response function is set to be 150~eV. The Tamm-Dancoff approximation was used to solve BSE. The highest one hundred valence bands and one hundred lowest conduction bands are considered as the basis for the excited state in the BSE procedure. The calculations are based on the optimized HSE06 functional which resulted in the optimized geometry and the electronic structure of the neutral tri-carbon interstitial cluster in 4H-SiC.


\section*{Author contribution}
PL carried out the calculations. All authors contributed to the discussion and writing the manuscript. AG developed the concept of electrically inactive defect emitters and led the entire scientific project.

\section*{Competing interests}
The authors declare that there are no competing interests.

\section*{Data Availability}
The data that support the findings of this study are available from the corresponding author upon reasonable request.

\section*{Acknowledgement}
Support by the National Excellence Program for the project of Quantum-coherent materials (NKFIH Grant No.\ KKP129866) as well as by the Ministry of Culture and Innovation and the National Research, Development and Innovation Office within the Quantum Information National Laboratory of Hungary (Grant No.\ 2022-2.1.1-NL-2022-00004) is much appreciated. AG acknowledges the high-performance computational resources provided by KIFÜ (Governmental Agency for IT Development) institute of Hungary and the European Commission for the project QuMicro (Grant No.\ 101046911). BH acknowledges the NSFC (Grants Nos.\ 12088101 and 12174404), National Key Research and Development of China (Grant No.\ 2022YFA1402400), NSAF (Grant No.\ U2230402).

\bibliography{ref}

\end{document}